\documentclass[twocolumn,showpacs,amsmath,amssymb]{revtex4}

\usepackage{graphics}
\usepackage{epsfig}

\begin{document}

\title{Measurement of the Positive Muon Anomalous Magnetic Moment to 0.7\,ppm}
\author{
G.W.~Bennett$^{2}$,
B.~Bousquet$^{9}$,
H.N.~Brown$^2$,
G.~Bunce$^2$,
R.M.~Carey$^1$,
P.~Cushman$^{9}$,
G.T.~Danby$^2$,
P.T.~Debevec$^7$,
M.~Deile$^{11}$,
H.~Deng$^{11}$,
W.~Deninger$^7$,
S.K.~Dhawan$^{11}$,
V.P.~Druzhinin$^3$,
L.~Duong$^{9}$,
E.~Efstathiadis$^1$,
F.J.M.~Farley$^{11}$,
G.V.~Fedotovich$^3$,
S.~Giron$^{9}$,
F.E.~Gray$^7$,
D.~Grigoriev$^3$,
M.~Grosse-Perdekamp$^{11}$,
A.~Grossmann$^6$,
M.F.~Hare$^1$,
D.W.~Hertzog$^7$,
X.~Huang$^1$,
V.W.~Hughes$^{11}$,
M.~Iwasaki$^{10}$,
K.~Jungmann$^5$,
D.~Kawall$^{11}$,
B.I.~Khazin$^3$,
J.~Kindem$^{9}$,
F.~Krienen$^1$,
I.~Kronkvist$^{9}$,
A.~Lam$^1$,
R.~Larsen$^2$,
Y.Y.~Lee$^2$,
I.~Logashenko$^{1,3}$,
R.~McNabb$^{9}$,
W.~Meng$^2$,
J.~Mi$^2$,
J.P.~Miller$^1$,
W.M.~Morse$^2$,  
D.~Nikas$^2$, 
C.J.G.~Onderwater$^7$,
Y.~Orlov$^4$,
C.S.~\"{O}zben$^2$,
J.M.~Paley$^1$,   
Q.~Peng$^1$,   
C.C.~Polly$^7$,   
J.~Pretz$^{11}$, 
R.~Prigl$^{2}$,
G.~zu~Putlitz$^6$,
T.~Qian$^{9}$,  
S.I.~Redin$^{3,11}$,
O.~Rind$^1$,
B.L.~Roberts$^1$,
N.~Ryskulov$^3$,
P.~Shagin$^9$,
Y.K.~Semertzidis$^2$, 
Yu.M.~Shatunov$^3$,
E.P.~Sichtermann$^{11}$,
E.~Solodov$^3$,
M.~Sossong$^7$, 
A.~Steinmetz$^{11}$,
L.R.~Sulak$^{1}$,
A.~Trofimov$^1$,
D.~Urner$^7$,
P.~von~Walter$^6$,
D.~Warburton$^2$,
and
A.~Yamamoto$^8$.
\\
(Muon $(g-2)$ Collaboration)
}

\affiliation{
\mbox{$\,^1$Department of Physics, Boston University, Boston, Massachusetts 02215}\\
\mbox{$\,^2$Brookhaven National Laboratory, Upton, New York 11973}\\
\mbox{$\,^3$Budker Institute of Nuclear Physics, Novosibirsk, Russia}\\
\mbox{$\,^4$Newman Laboratory, Cornell University, Ithaca, New York 14853}\\
\mbox{$\,^5$ Kernfysisch Versneller Instituut, Rijksuniversiteit Groningen, NL 9747\,AA Groningen, The Netherlands}\\
\mbox{$\,^6$ Physikalisches Institut der Universit\"at Heidelberg, 69120 Heidelberg, Germany}\\
\mbox{$\,^7$ Department of Physics, University of Illinois at Urbana-Champaign, Illinois 61801}\\
\mbox{$\,^8$ KEK, High Energy Accelerator Research Organization, Tsukuba, Ibaraki 305-0801, Japan}\\
\mbox{$\,^{9}$Department of Physics, University of Minnesota,Minneapolis, Minnesota 55455}\\
\mbox{$\,^{10}$ Tokyo Institute of Technology, Tokyo, Japan}\\
\mbox{$\,^{11}$ Department of Physics, Yale University, New Haven, Connecticut 06520}
}

\begin{abstract}
A higher precision measurement of the anomalous $g$ value, $a_\mu = (g-2)/2$, for the positive muon has been made at the Brookhaven Alternating Gradient Synchrotron, based on data collected in the year 2000.
The result $a_{\mu^+} = 11\,659\,204(7)(5) \times 10^{-10}$ (0.7\,ppm) is in good agreement with previous measurements and has an error about one half that of the combined previous data.
The present world average experimental value is $a_\mu(\mathrm{exp}) = 11\,659\,203(8) \times 10^{-10}$ (0.7\,ppm).
\end{abstract}

\maketitle
The study of magnetic moments has played an important role in our understanding of sub-atomic physics.
Precision measurements of the electron anomalous magnetic moment, together with those of the hyperfine structure of hydrogen and the Lamb shift, stimulated the development of modern quantum electrodynamics and have since provided stringent tests of this theory.
In this Letter we report a new measurement of the anomalous magnetic moment of the positive muon, $a_{\mu}=(g-2)/2$, with a relative precision of 0.7 parts per million (ppm), nearly two times better than our previous work \cite{Carey:1999dd,Brown:2000sj,Brown:2001mg}.
This measurement comes from data collected in the year 2000.
At this level, $a_{\mu}$ is sensitive to QED, weak, and hadronic virtual loops and
provides an important constraint on extensions to the standard model.

The principle of the experiment and previous results have been given in earlier publications~\cite{Carey:1999dd,Brown:2000sj,Brown:2001mg}.
Also, detailed descriptions of the $(g-2)$ superconducting inflector magnet, storage ring magnet, fast kicker, NMR system, and calorimeters have been published~\cite{Danby:2001eh}.

The quantity $a_\mu$ is determined from
\begin{equation}
  a_\mu = \frac{\omega_a}{\frac{e}{m_\mu c} \langle B \rangle}.
  \label{eq:amu}
\end{equation}
The  magnetic field $\langle B \rangle$ weighted over the muon beam distribution is measured by proton NMR. 
The difference frequency $\omega_a$ between the muon spin precession and orbital angular frequencies is determined by counting the number $N(t)$ of decay positrons with energies larger than an energy threshold,
\begin{equation}
  N(t) = N_0(E) e^{-t/(\gamma\tau)} \left[ 1 + A(E) \sin(\omega_a t + \phi_a(E)) \right].
  \label{eq:spectrum}
\end{equation}
The normalization $N_0$, asymmetry $A$, and phase $\phi_a$ vary with the chosen threshold.
The time dilated lifetime is \mbox{$\gamma\tau \approx 64.4\,\mu\mathrm{s}$.}
For muons with $\gamma = 29.3$, the angular difference frequency $\omega_a$ is not affected by electrostatic focusing fields in the ring.

New aspects of the 2000 data taking period include: the operation of the AGS with 12 beam bunches, contributing to a 4-fold increase in data collected as compared to 1999; a new  superconducting inflector magnet, which improved the field homogeneity in the muon storage region; the installation and operation of a sweeper magnet in the beamline, which reduced AGS background; and additional muon loss detectors, which enable an improved determination of the time dependence of muon losses.
Most other experimental aspects of the data taking in 2000 were the same as in 1998 and 1999.

\begin{figure}
  \includegraphics[width=0.45\textwidth,angle=0]{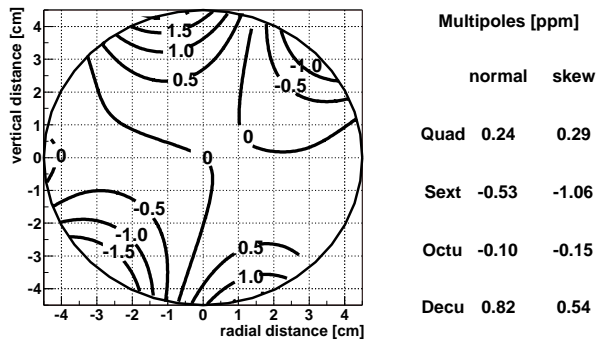}
  \caption{A 2-dimensional multipole expansion of the field averaged over azimuth from one out of 22 trolley measurements.  Half ppm contours  with respect to a central azimuthal average field \mbox{$B_0 = 1.451\,274\,\mathrm{T}$} are shown.  The multipole amplitudes relative to $B_0$, are given at the beam aperture, which has a  radius of 4.5\,cm and is indicated by the circle.}
  \label{fig:field}
\end{figure}

The magnetic field value was obtained from NMR measurements of the free proton resonance frequency.
A trolley with 17 NMR probes was used to measure the field throughout the muon storage ring, typically every three days.
The trolley probes were calibrated to an accuracy of 0.15\,ppm with respect to a standard spherical \mbox{$\mathrm{H}_2\mathrm{O}$} probe, which has an absolute calibration known to 0.05\,ppm.
The probes were intercalibrated during the data taking period using a single, movable probe plunged into the storage region.
Interpolation of the magnetic field between trolley measurements was based on the continual readings of about 150 fixed NMR probes distributed around the ring in the top and bottom walls of the vacuum chamber.
Figure~\ref{fig:field} shows a magnetic field profile averaged over azimuth.
The improved field homogeneity obtained with the new inflector relaxes the demands on the knowledge of the muon beam distribution.

The field $\langle B \rangle$ averaged over the muon beam distribution in space and time was obtained from two complete and largely independent analyses, whose results were found to agree to within 0.05\,ppm.
Its final value is expressed in terms of the free proton resonance frequency and is given by
\mbox{$\omega_p/(2\pi) = 61\,791\,595(15)\,\mathrm{Hz}$ (0.2\,ppm).}
Table~\ref{table:field} lists the uncertainties.
\newsavebox{\tempbox}
\sbox{\tempbox}{
  \begin{minipage}{0.45\textwidth}
    \begin{small}
      $^\dagger$ higher multipoles, trolley temperature and voltage response,
      eddy currents from the kickers, and time-varying stray fields.
    \end{small}
  \end{minipage}
}
\begin{table}
\caption {Systematic uncertainties for the $\omega_p$ analysis}
\begin{tabular}{l|c}
\hline\hline
Source of errors & Size [ppm] \\
\hline
Absolute calibration of standard probe\hspace{3em} & 0.05\\
Calibration of trolley probe & 0.15\\
Trolley measurements of $B_0$ & 0.10\\
Interpolation with fixed probes & 0.10\\
Uncertainty from muon distribution & 0.03\\
Others$^\dagger$ & 0.10\\
\hline
Total systematic error on $\omega_p$ & 0.24 \\
\hline\hline
\end{tabular}
\usebox{\tempbox}
\label{table:field}
\end{table}

The frequency $\omega_a$ was obtained from the time distribution of decay positron counts.
The positrons were detected with 24 lead/scintillating fiber calorimeters on the inside of the storage ring, whose photomultiplier signals had a typical FWHM of 5\,ns.
The signals were recorded with 400\,MHz, 8-bit waveform digitizers (WFD).
Pulses above a preset energy threshold of about 1\,GeV triggered the WFD to record at least 16 ADC samples (40\,ns).
Pulses with energies below the hardware threshold were recorded if they appeared within the sampling window around a trigger pulse.
The positron arrival times and energies were reconstructed offline from the WFD recordings using two independent implementations of our pulse reconstruction algorithm.

The large positron sample and 30\% higher instantaneous count rates than in 1999 required careful consideration of distortions to the spectrum in Eq.~\ref{eq:spectrum}.
As in the analysis of our 1999 data, we have considered (1) positron pulses overlapping in time (pileup), (2) coherent betatron oscillations, (3) beam debunching, (4) muon losses, and (5) detector gain and time instability.

(1) Pileup of positron signals distorts the time spectrum because of misidentification of the number, energies, and times of the positrons.
The pileup spectrum can be constructed from the data using the extended pulse sampling by the WFD described above, and can effectively be subtracted from the data prior to fitting~\cite{Brown:2001mg}.

(2) As in earlier data taking periods, the weak focusing storage ring was operated with a field index $n = 0.137$, well away from beam and spin resonances.
The phase space for betatron oscillations defined by the acceptance of the storage ring was not filled, which resulted in coherent betatron oscillations (CBO) --- betatron oscillations of the beam as a whole.
Since the acceptance of a calorimeter varies with the radial muon decay position in the storage ring and the momentum of the produced decay positron, the observed positron time and energy spectra are modulated with the CBO frequencies.
The most important modulation is the horizontal one with frequency
 \mbox{$\omega_\mathrm{cbo,h} 
= \left(1-\sqrt{1-n}\,\right) \omega_c = 2\pi\cdot466$\,kHz},
where $\omega_c$ is the cyclotron frequency.
At injection time, this modulation affects the terms $N_0$, $A$, and $\phi_a$ in Eq.~\ref{eq:spectrum} at the level of 1\%, 0.1\%, and 1\,mrad, respectively.

For $n = 0.137$, the frequency $\omega_\mathrm{cbo,h}$ is approximately twice as large as $\omega_a$ and, hence, the interference frequency $\omega_\mathrm{cbo,h}-\omega_a$ is close to $\omega_a$.
Modulations of the asymmetry and phase with frequencies $\omega_\mathrm{cbo,h} \simeq 2\times\omega_a$ may, unlike the modulation of observed counts, manifest themselves as sizable, artificial shifts in the fitted frequency value $\omega_a$, when not taken into account in the function fitted to the data.
These shifts are as large as 4\,ppm for individual calorimeter time spectra, and mostly cancel in the summed spectra owing to the circular symmetry of the experiment design.

\begin{figure}
  \includegraphics[width=0.45\textwidth,angle=0]{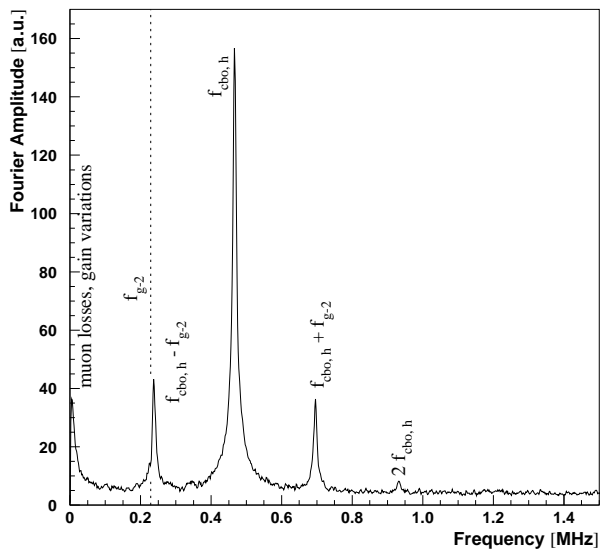}
 \caption{Coherent betatron oscillations (CBO) in the $g-2$ time spectra.  The Fourier spectrum was obtained from residuals from a fit based on muon decay and spin precession (Eq.~\ref{eq:spectrum}) alone.  The horizontal modulation was at $\omega_\mathrm{cbo,h}/(2\pi)$ = 466\,kHz in the year 2000, so that the interference frequency $\omega_\mathrm{cbo,h}-\omega_a$ is numerically close to $\omega_a$, as indicated.  The frequency $\omega_a$ is determined from fits that take CBO into account.}
  \label{fig:fourier}
\end{figure}

The existence of CBO and their effect on the positron time spectra can be seen from Fig.~\ref{fig:fourier}, showing the Fourier amplitudes of residuals from fits to the data based only on muon decay and spin precession (Eq.~\ref{eq:spectrum}).
The Fourier amplitudes at the interference frequencies $\omega_\mathrm{cbo,h} \pm \omega_a$ differ by about 20\% because of the aforementioned modulation of the observed phase.
The characteristic CBO decay times were $\sim 100\,\mu$s.

(3) The  injection of the beam in narrow bunches into the storage ring resulted in a  strong  modulation  of  the  initial  part  of  the  positron time spectrum with the 149.2\,ns cyclotron period.
This effect was removed from the analyses by uniformly randomizing the recorded start time for each beam pulse over one cyclotron period.

(4) Losses of muons during the data collection were minimized by controlled scraping~\cite{Brown:2000sj} of the beam before the data collection started.
The time dependence of small, residual losses was determined by forming triple coincidences from scintillating detectors mounted to the face of three adjacent calorimeters.
Corrections were made for accidental coincidences and proton background.
The time spectrum of lost muons was measured, up to a normalization, at five locations in the storage ring.

(5) Average calorimeter timing shifts were determined with a pulsed 
UV laser system to be typically smaller than 4\,ps over the 
first \mbox{200\,$\mu$s} of data taking.
Detector gain changes are determined from the average observed positron energy as a function of time after beam injection.
>From 50\,$\mu$s on, 5\,$\mu$s after the last of the calorimeters was gated on in 2000, the gains of all but 2 of the calorimeters were stable to within 0.2\% over the 10 dilated muon lifetimes of data collection.

The event sample collected in the year 2000 amounts, after data selection, to $4\cdot10^9$ positrons with energies greater than 2\,GeV in the time region \mbox{50\,$\mu$s} to \mbox{600\,$\mu$s} following muon injection into the storage ring.
A statistical uncertainty in the fitted frequency $\omega_a$ of about 0.7\,ppm results.
To assess systematic uncertainties to an adequate precision, distortions of the decay positron spectrum in Eq.~\ref{eq:spectrum} have been studied in the full range of observed energies.
In addition, the time spectrum for positron energies exceeding 2\,GeV has been studied, as well as two representations of the time spectra that suppress periodic distortions with frequency $\omega_\mathrm{cbo,h}$ and slow distortions, respectively.
In total, four independent and complete analyses of $\omega_a$ were performed.

In the first analysis the data were fitted in 0.2\,GeV energy intervals in the range 1.4--3.2\,GeV for each detector separately.
This results in 198 independent fits.
The pileup contribution to each spectrum was determined from the WFD recordings using a variant of the technique used before.
For each energy interval and each calorimeter station, pileup was statistically constructed and fitted to an appropriate functional form.
The resulting parameters were then incorporated into the fitting function used to describe the positron time spectra.
Horizontal CBO were incorporated in the function fitted to the data through modulation of  the observed number of counts, asymmetry, and phase.
The function fitted to the calorimeter data in each energy interval is given by
\begin{equation}
  N(t) = N_0(t;E) e^{-t/(\gamma\tau)} \left[ 1 + A(t;E) \sin(\omega_a t + \phi_a(t;E)) \right],
  \label{eq:physicsfunction}
\end{equation}
in which the time dependences of $N_0$, $A$, and $\phi_a$ are all modulated as
$\sim 1 + \mathcal{A}_i(t)\sin(\omega_\mathrm{cbo,h} t + \phi_i)$,
where $i$ stands for any of the three separate modulations of $N_0$, $A$, and $\phi_a$.
The modulation amplitudes were determined empirically from the data and follow primarily an exponential with a characteristic decay time of approximately \mbox{100\,$\mu$s}.
Muon losses were constrained to the shape of the measured losses. 
The spectrum described by Eq.~\ref{eq:physicsfunction} is multiplied by a time-dependent factor which accounts for these losses.
Fit start times for each calorimeter station and energy interval were chosen at  times after which the quality of fit has become constant.
Slowly vanishing distortions of the exponential, due to imperfectly corrected energy scale variations, pileup, or muon losses, are thereby removed.

In the second analysis the time spectrum of all positrons with energies larger than 2\,GeV and arrival times in the region 49\,$\mu$s to 600\,$\mu$s after muon injection was considered.
The pileup contribution to the spectrum was determined from the data and corrected for as in Ref.~\cite{Brown:2001mg}.
The effects of horizontal CBO were incorporated
as modulations of the observed number of counts and asymmetry, whereas the modulation of the observed phase was neglected in the fitting function.
The CBO modulation envelope was determined by partial Fourier integration and confirmed from the modulation of the observed average energies for equal phases of the $(g-2)$ oscillation.
The time dilated muon lifetime $\gamma\tau$ was fixed to the expected value.
The treatment of muon losses followed the one described above.
Slowly vanishing distortions of the positron spectra were parametrized empirically, and were included in the fitting function to improve the overall quality of fit.
Their correlation with the fitted frequency $\omega_a$ is weak.

In the third analysis, as in the second analysis, $\omega_a$ was determined from a multiparameter fit to the summed time spectrum of positrons detected with all calorimeters.
Alternative methods were used to assess systematic uncertainties.
In particular, to study the systematics associated with CBO, the data were strobed at fixed phases of the horizontal CBO modulation.
Since $\omega_\mathrm{cbo,h} > 2 \times \omega_a$, the $(g-2)$ signal can be fully determined from the resulting spectra.
Eq.~\ref{eq:spectrum} was fitted to the spectra after strobing, giving a result for $\omega_a$ which is less sensitive by a factor of about three to CBO modulations of the asymmetry and phase than results from fits to the unstrobed spectrum.

In the fourth analysis the ratio introduced in Eq.~6 of Ref.~\cite{Brown:2001mg},
$r(t) \sim A(E)\sin(\omega_a t + \phi_a(E))$, was fitted.
The ratio is largely insensitive to changes of observed counts on time scales larger than $\tau_a = 2\pi/\omega_a \sim 4\,\mu\mathrm{s}$.
The effects of asymmetry and phase modulation were not explicitly included in the fitting function.
This analysis used summed calorimeter spectra, in which the asymmetry and phase modulation effects are suppressed by an order of magnitude as compared to summed results from fits to individual calorimeter spectra.
The suppression originates in the symmetry of the experiment design, and has been studied extensively by including the effects in the fits, by determining the energy modulation, and by Monte Carlo simulations.

\begin{figure}
  \includegraphics[width=0.45\textwidth,angle=0]{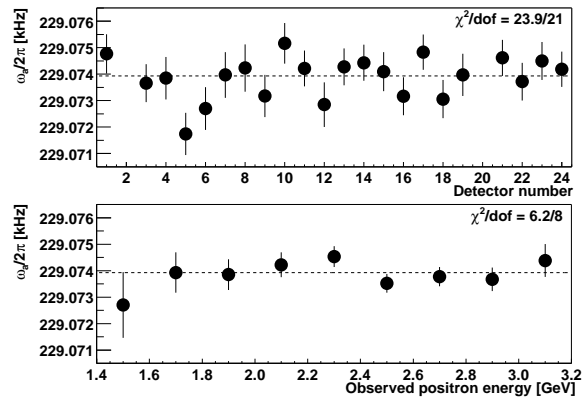}
  \caption{a) The frequency $\omega_a/(2\pi)$ determined from fits to the individual calorimeter time spectra.  Data from calorimeters 2 and 20 were discarded, as in the analysis of our 1999 data.
 b) The fitted frequency $\omega_a/(2\pi)$ versus positron energy.
 These results come from the analysis described first in the text.
 }
  \label{fig:consistency}
\end{figure}

The internal consistency of the results was verified in various ways.
For all analyses the fitted frequency $\omega_a$ was found constant with the fit start-time, up to statistical fluctuations.
The Fourier transform of residuals from fits incorporating asymmetry and phase modulation exhibits none of the CBO related interference structure in Fig~\ref{fig:fourier}.
Fig.~\ref{fig:consistency}a shows the results for fits to the spectra from individual detectors from the analysis described first.
Fig.~\ref{fig:consistency}b shows the results on $\omega_a$ versus positron energy.

The results from the analyses are found to agree, on $\omega_a$ to within 0.4\,ppm.
This is within the statistical variation of 0.5\,ppm expected from the use of slightly different data selections and the different treatment of the data in the respective analyses.
Each of the analyses gives results with comparable total uncertainties in $\omega_a$.
The analysis which incorporates asymmetry and phase modulation in the 
fitted function results in the smallest systematic uncertainty, primarily 
since CBO related systematic uncertainties are at the level of 0.05\,ppm 
in this approach.
The trade-off is a slightly enlarged statistical uncertainty, since more 
CBO related parameters are fitted.
The systematic uncertainty due to CBO is larger than our 
1999 estimate~\cite{Brown:2001mg} which
did not include the effects from asymmetry 
and phase modulation. 
Neither the 1999 value of $a_{\mu}$, nor its total uncertainty, were
changed by this improved estimate.
Other uncertainties are comparable or smaller.

We combine the present results to $\omega_a/(2\pi) = 229\,074.11(14)(7)\,\mathrm{Hz}$ (0.7\,ppm), which includes a correction of +0.76(3)\,ppm for contributions to Eq.~\ref{eq:amu} caused by vertical oscillations and, for muons with \mbox{$\gamma \neq 29.3$}, by horizontal electric fields.
The stated uncertainties account for strong correlations among the individual results, both statistical and systematic.
Table~\ref{table:omegaa} lists the  systematic uncertainties in the combined result with these correlations taken into account.
\sbox{\tempbox}{
  \begin{minipage}{0.45\textwidth}
    \begin{small}
    \noindent$^\dagger$ AGS background, timing shifts, E field and vertical oscillations,
                                           beam debunching/randomization.
    \end{small}
  \end{minipage}
}
\begin{table}
\caption{Systematic uncertainties for the $\omega_a$ analysis.}
\begin{tabular}{l|c}
\hline\hline
Source of errors & Size [ppm] \\
\hline
Coherent betatron oscillations\hspace{6em}  &  0.21\\
Pileup & 0.13\\
Gain changes & 0.13\\
Lost muons & 0.10\\
Binning and fitting procedure & 0.06\\
Others$^\dagger$ & 0.06\\
\hline
Total systematic error on $\omega_a$ & 0.31\\
\hline\hline
\end{tabular}
\usebox{\tempbox}
\label{table:omegaa}
\end{table}

After the $\omega_p$ and $\omega_a$ analyses were finalized separately and independently, $a_\mu$ was evaluated.
The result is
\begin{equation}
  a_{\mu^+} = \frac{R}{\lambda - R} = 11\,659\,204(7)(5)\ \times\ 10^{-10}~~\mbox{(0.7\,ppm)},
  \label{eq:result}
\end{equation}
in which $R = \omega_a/\omega_p$ and $\lambda = \mu_\mu/\mu_p = 3.183\,345\,39(10)$~\cite{Groom:2000in}.
This new result is in good agreement with previous measurements~\cite{Bailey:1979,Carey:1999dd,Brown:2000sj,Brown:2001mg}, and reduces the combined uncertainty by about half.
The present world average experimental value is
\begin{equation}
  a_\mu(\mathrm{exp}) = 11\,659\,203(8) \times 10^{-10}~~\mbox{(0.7\,ppm)},
\end{equation}
which is driven by our determinations of
$a_\mu$ in Refs.~\cite{Brown:2000sj, Brown:2001mg} and Eq.~\ref{eq:result},
and accounts for correlations between systematic uncertainties.

The theoretical value of $a_\mu$ in the standard model (SM) is determined from $a_\mu(\mathrm{SM}) = a_\mu(\mathrm{QED}) + a_\mu(\mathrm{had}) + a_\mu(\mathrm{weak})$,
in which
$a_\mu(\mathrm{QED}) = 11\,658\,470.57(0.29)\,\times\,10^{-10}$ (0.025\,ppm) \cite{Mohr:2000} and
$a_\mu(\mathrm{weak}) = 15.1(0.4)\,\times\,10^{-10}$ (0.03\,ppm) \cite{Czarnecki:2001pv}.
The hadronic contribution $a_\mu(\mathrm{had})$ receives its leading
contribution from $a_\mu(\mathrm{had,1})$, which is currently evaluated to
be $a_\mu(\mathrm{had,1}) = 692(6)\,\times\,10^{-10}$
(0.6\,ppm)~\cite{Davier:1998si}.
The two more recent, published evaluations~\cite{Narison:2001jt,DeTroconiz:2001wt} used recent, preliminary data from Novosibirsk which have since been superseded~\cite{Akhmetshin:2001ig}, so we choose not to use their values for comparison.
Higher order contributions include
$a_\mu(\mathrm{had,2}) = -10.0(0.6)\,\times\,10^{-10}$ \cite{Krause:1997rf}
and the contribution from hadronic light-by-light scattering, which we now take to be $a_\mu(\mathrm{had,lbl}) = 8.6(3.2)\,\times\,10^{-10}$
 \cite{Knecht:2001qg}.
The chiral perturbation calculation of Ref.~\cite{Ramsey-Musolf:2002cy} provides an estimate with larger uncertainty.
Hence, the value of $a_\mu(\mathrm{SM})$ is currently evaluated to be,
\begin{equation}
  a_\mu(\mathrm{SM}) = 11\,659\,177(7)\,\times\,10^{-10} \mathrm{\ (0.6\,ppm)}
\end{equation}
Additional data on $e^+e^-$ collisions~\cite{Akhmetshin:2001ig,Zhao:2000xc} and on $\tau$-decay~\cite{Anderson:1999ui} have been published, and are being considered in future evaluations of $a_\mu(\mathrm{had,1})$.
New data can be expected from the Frascati $\phi$ factory and from the $B$ factories.

\begin{figure}
  \includegraphics[width=0.43\textwidth,angle=0]{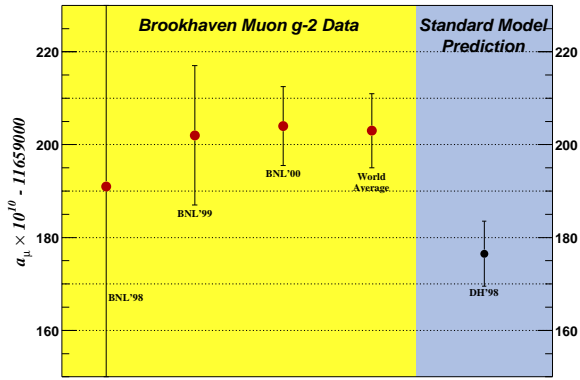}
  \caption{Recent measurements of $a_\mu$, together with the standard model prediction using the evaluation in Ref.~\cite{Davier:1998si} of $a_\mu(\mathrm{had,1})$ from $e^+e^-$ and $\tau$ decay data.
}
  \label{fig:results}
\end{figure}

The three most recent measurements of $a_\mu$ along with the above standard model prediction are shown in  Fig.~\ref{fig:results}.
The present experimental uncertainty is about half the size of the weak contribution to $a_\mu(\mathrm{SM})$.
The difference of $a_\mu(\mathrm{exp})$ and $a_\mu(\mathrm{SM})$ is about 2.6 times the combined experimental and stated theoretical uncertainty.
 
In 2001 data on $\mu^-$ were obtained.
Approximately \mbox{$3 \times 10^9$} decay electrons were observed.
Field focusing indices $n = 0.122$ and $n = 0.142$ were used.
Measurement of $a_{\mu^-}$ will provide a sensitive test of CPT violation and also an improved value of $a_\mu$.
We plan further data-taking with $\mu^-$ to obtain an additional \mbox{$6 \times 10^9$}  counts.

We thank T.~Kirk, D.I.~Lowenstein, P.~Pile, and the staff of the BNL AGS for the strong support they have given this experiment.  We thank J.~Bijnens, A.~Czarnecki, M.~Davier, S.I.~Eidelman, A.~H\"{o}cker, F.~Jegerlehner, T.~Kinoshita, W.~Marciano, and E.~de~Rafael for helpful discussions.
This work was supported in part by the U.S. Department of Energy, the U.S. National Science Foundation, the U.S. National Computational Science Alliance, the German Bundesminister f\"{u}r Bildung und Forschung, the Russian Ministry of Science, and the U.S.-Japan Agreement in High Energy Physics.
M.~Deile acknowledges partial support by the Alexander von Humboldt Foundation.
F.E.~Gray was supported in part by a General Electric Fellowship,
and C.S.~\"{O}zben by the Scientific and Research Council of Turkey (TUBITAK).



\end{document}